\documentclass[twocolumn,prl,showpacs]{revtex4}

\usepackage[pdftex]{graphicx,color}
\usepackage{amsmath}
\usepackage{amssymb}
\usepackage{natbib}
\usepackage{dcolumn}

\begin{document}

\title{High-fidelity transport of trapped-ion qubits through an X-junction
trap array}

\author{R.B. Blakestad}
\email{blakesta@boulder.nist.gov}
\author{C. Ospelkaus}
\author{A.P. VanDevender}
\author{J.M. Amini}
\author{J. Britton}
\author{D. Leibfried}
\author{D.J. Wineland}

\affiliation{National Institute of Standards and Technology, 325 Broadway, Boulder,
Colorado 80305, USA }

\date{March 30, 2009}

\begin{abstract}
We report reliable transport of $^9$Be$^+$ ions through a 2-D trap array that
includes a separate loading/reservoir zone and an ``X-junction".  During
transport the ion's kinetic energy in its local well increases by only a few
motional quanta and internal-state coherences are preserved. We also examine two
sources of energy gain during transport: a particular radio-frequency (RF) noise
heating mechanism and digital sampling noise.  Such studies are important to
achieve scaling in a trapped-ion quantum information processor.
\end{abstract}
\maketitle

Key requirements for efficient large-scale quantum information processing (QIP)
include reliable transport of information throughout the processor and the
ability to perform gates between arbitrary qubits. Trapped ions are a useful
system for studying elements of QIP~\cite{blatt2008a,monroe2008a,haffner2008a}
and can potentially satisfy these requirements. The multi-qubit gates that have
been demonstrated couple ions through a shared mode of
motion~\cite{blatt2008a,monroe2008a,haffner2008a}, similar to the proposal
in~\cite{cirac1995a}. This mode should be cooled to the ground state for maximum
fidelity, which has exceeded 0.99~\cite{benhelm2008a} in demonstrations. However,
as the number of ions $N$ grows, it becomes increasingly difficult to address the
desired mode without coupling to other modes. To achieve scalability, the ions
could be distributed over separate trap zones where $N$ for each zone remains
small~\cite{bible,cirac2000a,kielpinski2002a}. Information could be shared
between zones by moving the ions~\cite{bible,cirac2000a,kielpinski2002a} or
connecting them with photons~\cite{devoe1998a,moehring2007b,monroe2008a}. In the
proposal of~\cite{bible,kielpinski2002a}, multidimensional arrays incorporating
junctions would enable ions selected from arbitrary locations to be grouped
together for multi-qubit gates. Ion transport between zones must be accomplished
with high success probability and minimal kinetic energy gain to not adversely
affect the fidelity of subsequent multi-qubit
gates~\cite{blatt2008a,monroe2008a,haffner2008a}. Sympathetic cooling can restore
the motional modes for gate operations~\cite{bible,kielpinski2002a}, but for
large kinetic energy, this costs significant time, accompanied by decoherence,
and possible ion reordering.

Previously, ions have been transported in a linear
array~\cite{rowe2002a,barrett2004a,hensinger2006a,stick2006a,huber2008a} and
separated for further processing without loss of qubit
coherence~\cite{rowe2002a,barrett2004a} and with minimal energy gain. Transport
through a two-dimensional ``T-junction" has been achieved with Cd$^+$
ions~\cite{hensinger2006a}. However, in that experiment, the success probability
for round-trip transit was~0.98, and ions experienced considerable energy gain,
estimated to be $\sim$1~eV, which would necessitate a substantial delay for
recooling before performing subsequent gates. In this letter, we report
transport of $^9$Be$^+$ ions through a junction with greater than
$0.9999$~probability and less than $10^{-7}$~eV energy gain.

\begin{figure}
\includegraphics[width=0.5\textwidth]{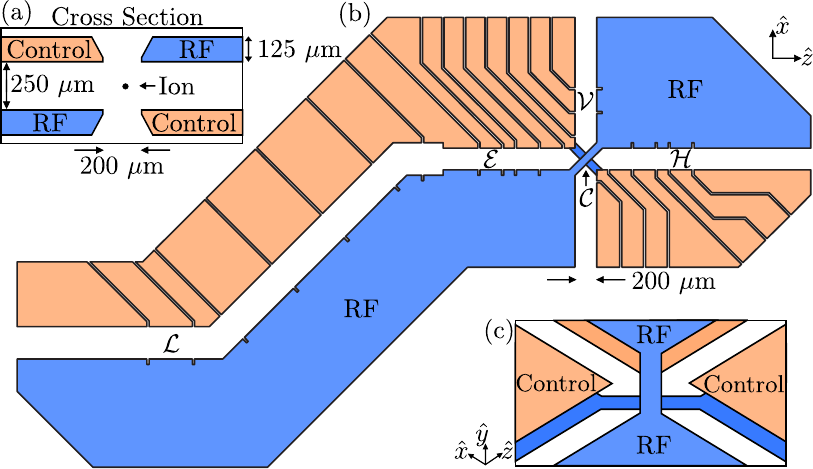}
\caption{\label{fig:trap_fig} (a) Cross-sectional view of the two layers of
electrodes. (b)  Top view of the electrode layout, with the RF electrodes
indicated, and all other (control) electrodes held at RF ground.  A nearly
identical set of electrodes sits below these electrodes, with RF and control
electrodes interchanged~\cite{rowe2002a}. Forty-six control electrodes support 18
different trapping zones. The load zone ($\mathcal{L}$), the main experiment zone
($\mathcal{E}$), the vertical zone ($\mathcal{V}$), the horizontal zone
($\mathcal{H}$) and the center of the junction  ($\mathcal{C}$) are labeled. (c)
Schematic of the RF bridges from an oblique angle (not to scale).
}
\end{figure}

The trap (Fig.~\ref{fig:trap_fig}), constructed from two laser-machined alumina
wafers with gold-coated electrodes~\cite{rowe2002a}, is derived from of a linear
RF Paul trap. It was held in vacuum at $< 5 \times 10^{-9}$~Pa, while a 1.44~mT
$B$-field along $\hat{y} - \hat{z}$ (the Doppler cooling and detection laser beam
direction) provided the ion internal-state quantization axis. Radial
confinement (perpendicular to the channel axes) was provided by an RF
($V_{0} \approx 200$~V, $\Omega_{\textrm{RF}} / 2 \pi \approx 83$~MHz)
pseudopotential with axial confinement due primarily to quasi-static control
potentials~\cite{rowe2002a}. Two main channels intersect at zone $\mathcal{C}$
(center), forming an ``X-junction''.  In addition to $\mathcal{C}$, zones
$\mathcal{E}$ (experiment), $\mathcal{H}$ (horizontal) and $\mathcal{V}$
(vertical), located in other legs of the junction, comprise the four destinations
for transport. The lower leg of the X-junction provides symmetry about
$\mathcal{C}$ but lacks the required electrodes to separately hold ions.
Typically, the imaging optics and laser beams were positioned at~$\mathcal{E}$,
though $\mathcal{L}$, $\mathcal{H}$ and $\mathcal{V}$ allow similar access. Laser
beam parameters were similar to those in~\cite{rowe2002a,barrett2004a}.

\begin{figure}
\includegraphics[width=0.5\textwidth]{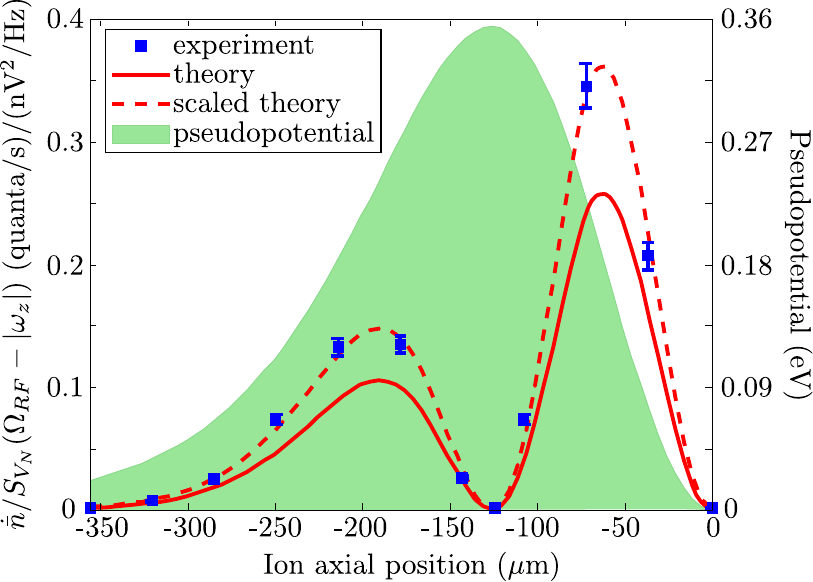}
\caption{\label{fig:rfheating} The ratio of heating rate $\dot{\bar{n}}$ to
voltage noise spectral density $S_{V_{\textrm{N}}} (\Omega_{\textrm{RF}} - \left|
\omega_z \right| )$ for various locations along the trap axis
($\mathcal{C}$ is located at 0~$\mu$m). The theoretical prediction is shown
both with and without a scaling parameter ($= 1.4$).  The simulated
pseudopotential is overlaid in the background, in units of eV.  Since heating
is gradient dependent, we see very little heating at the peak of the
pseudopotential barrier, despite it being the point of maximum (axial) RF
electric field and ion micromotion. Nearly identical pseudopotential barriers
occur on the other three legs of the junction.}
\end{figure}

Two diagonal bridges connect the RF electrodes across the junction
(Fig.~\ref{fig:trap_fig}c).  These bridges induce pseudopotential barriers
($\sim$0.35~eV) along the $\hat{x}$ and $\hat{z}$ channels, adjacent to the
junction (Fig.~\ref{fig:rfheating}), which must be overcome to transport ions
through the junction. Narrow control electrodes (100~$\mu$m wide) near the
junction accomplish this while using modest voltages ($|V| \leq 10$~V).

The time-varying potentials on the control electrodes were designed to produce a
moving harmonic well (axial frequency $\omega_z$) that adiabatically transported
ions through the trap array and minimized energy gain in the motional mode. This
appears to be a crucial difference from the experiments in~\cite{hensinger2006a}
where, when the ions were transported across a pseudopotential barrier, they
gained energy by falling from its apex.  In our experiments, $\omega_z / 2 \pi $
increased from 3.6~MHz at $\mathcal{E}$, $\mathcal{H}$, and $\mathcal{V}$ to
5.7~MHz as the ion neared $\mathcal{C}$, where symmetry forces $\omega_z$ to be
close to one of the two radial-mode frequencies.  Radial frequencies ranged from
12 to 14~MHz except near $\mathcal{C}$, where one decreased to 5.7~MHz.

Ions are created in zone~$\mathcal{L}$ by photo-ionizing neutral Be, which is
delivered from a heated source.  The electrodes of other zones are shielded from
Be deposition by a mask (not shown). Zone~$\mathcal{L}$ can be used as a
reservoir for ions, but in the experiments here, ions are directly shuttled
to~$\mathcal{E}$ where they are laser cooled. Following transport to
$\mathcal{C}$, $\mathcal{H}$, or $\mathcal{V}$, they were returned to
$\mathcal{E}$, where the average harmonic occupation $\bar{n}$ was measured for
the axial mode, yielding the motional energy gain. We report three cases:
transport to the junction's center and back
($\mathcal{E}$-$\mathcal{C}$-$\mathcal{E}$), transport through the junction
horizontally
($\mathcal{E}$-$\mathcal{C}$-$\mathcal{H}$-$\mathcal{C}$-$\mathcal{E}$), and
vertically
($\mathcal{E}$-$\mathcal{C}$-$\mathcal{V}$-$\mathcal{C}$-$\mathcal{E}$). By
symmetry, all paths leading away from $\mathcal{C}$ are equivalent. Therefore,
$\mathcal{E}$-$\mathcal{C}$-$\mathcal{E}$ transport and transporting through the
junction once (i.e., $\mathcal{E}$-$\mathcal{C}$-$\mathcal{H}$) are analogous. In
the adiabatic regime, for one ion we expect no additional energy gain due to
turning the corner toward $\mathcal{V}$.

Energy gain after transport was determined using two independent methods
(Table~\ref{tab:heatingrates}). The Doppler recooling
method~\cite{epstein2007b,wesenberg2007a} was sensitive to $\bar{n} \gtrsim 50$.
For energy gain below 50~quanta/trip, this method required performing multiple
transports before measurement. In the Rabi flopping
method~\cite{meekhof1996a,rowe2002a}, which was sensitive for $\bar{n} \lesssim
100$, flopping curves on the ``red'' and ``blue'' sidebands or ``carrier'' of a
stimulated-Raman-induced groundstate hyperfine transition were obtained. These
curves were fit using calibrated Rabi rates for different Fock states $ \left| n
\right> $ to extract~$\bar{n}$. Fits assuming an arbitrary state, a thermal
state, or a coherent state were made.  Each assumption gave a slightly different
result for $\bar{n}$ (range given in Table~\ref{tab:heatingrates}) with no
obvious preferred method. Nevertheless, the approximate agreement between
recooling and flopping estimates lends confidence to the overall results.

\begin{table}
\caption{\label{tab:heatingrates} Axial-motion energy gain for transport
through the X-junction, for one and two $^9$Be$^+$ ions. Increase in the average
motional state quantum number $\bar{n}$ was measured by two methods:
``Doppler recooling'' and ``Rabi flopping'' (see text). The energy gain is
stated in units of quanta in a 3.6~MHz trapping well where $\bar{n}=5$~quanta
corresponds to 82~neV. }
\begin{ruledtabular}
\begin{tabular}{lccc}
Transport&&\multicolumn{2}{c}{Energy Gain (quanta/trip)}\\
&&recooling&flopping\\
\hline
$\mathcal{E}$-$\mathcal{C}$-$\mathcal{E}$&1 ion
& $3.2 \pm 1.8$			& 5--12 \\

$\mathcal{E}$-$\mathcal{C}$-$\mathcal{H}$-$\mathcal{C}$-$\mathcal{E}$&1 ion
& $7.9 \pm 1.5$			        & 6--12 \\

$\mathcal{E}$-$\mathcal{C}$-$\mathcal{V}$-$\mathcal{C}$-$\mathcal{E}$&1 ion
& $14.5 \pm 2.0$			       & 7--14 \\

\hline
$\mathcal{E}$-$\mathcal{C}$-$\mathcal{E}$&2 ions
& $5.4 \pm 1.2$	  		      & --- \\

$\mathcal{E}$-$\mathcal{C}$-$\mathcal{H}$-$\mathcal{C}$-$\mathcal{E}$&2 ions
& $16.6 \pm 1.8$	       		& --- \\

$\mathcal{E}$-$\mathcal{C}$-$\mathcal{V}$-$\mathcal{C}$-$\mathcal{E}$&2 ions  
& $53.0 \pm 1.2$			       & --- \\
\end{tabular}
\end{ruledtabular}
\end{table}

The intrinsic heating rate for a stationary ion at~$\mathcal{E}$ was measured to
be 40~quanta/s, which negligibly affects the results. We expect that the dominant
kinetic energy increase will occur along the axis (the direction of transport),
which is the direction employed in demonstrated qubit gates. This is supported by
the approximate agreement of energy gains estimated using Rabi flopping
(sensitive primarily to axial energy) and Doppler recooling (sensitive to both
axial and transverse energy; but in the analysis, we attribute all heating to the
axial mode). Transport between $\mathcal{C}$ and $\mathcal{E}$ or $\mathcal{H}$
(890~$\mu$m) took approximately 50~$\mu$s, with 20~$\mu$s to cross the
pseudopotential barrier. Transport from~$\mathcal{E}$ to~$\mathcal{H}$
or~$\mathcal{V}$ took approximately 100~$\mu$s.

To determine transport success, 10,000
$\mathcal{E}$-$\mathcal{C}$-$\mathcal{H}$-$\mathcal{C}$-$\mathcal{E}$ transports
were performed while verifying that the ion always appeared in $\mathcal{E}$ as
intended.  Then 10,000 more transports were performed, now verifying that the ion
always appeared in $\mathcal{H}$. With no observed failures in either set, we
place a lower limit on the transport success probability of 0.9999.  The same
procedure was used to verify
$\mathcal{E}$-$\mathcal{C}$-$\mathcal{V}$-$\mathcal{C}$-$\mathcal{E}$ transport
with identical results.  Ion lifetime, and thus transport success probability,
would be ultimately limited by ion loss resulting form background-gas
collisions~\cite{bible}. With this in mind, we observed no increase in ion-loss
rates over that for a stationary ion ($\sim 0.5/$hr) during the transport
experiments.  By observing millions of successive round trips for the three types
of transport, this suggests a success probability of greater than 0.999995. Since
transport comprised a small fraction of the total experiment duration, many of
these losses likely occurred when the ion was not being transported.

Moving pairs of ions in the same trapping well would be useful for both
sympathetic cooling and efficient ion manipulation~\cite{kielpinski2002a}.  We
demonstrated such transport using pairs of $^9$Be$^+$ ions and observed loss
rates comparable to those for an ion pair held stationary. Additional heating
mechanisms for multiple ions~\cite{bible,walther1993b} may explain the higher
transport energy gain observed for the pair (Table~\ref{tab:heatingrates}).
Additional energy was likely gained during rotation of the ion pair at
$\mathcal{C}$ for the
$\mathcal{E}$-$\mathcal{C}$-$\mathcal{V}$-$\mathcal{C}$-$\mathcal{E}$ transport.
Single $^{24}$Mg$^+$ ions have also been successfully transported.

\begin{figure}
\includegraphics[width=0.5\textwidth]{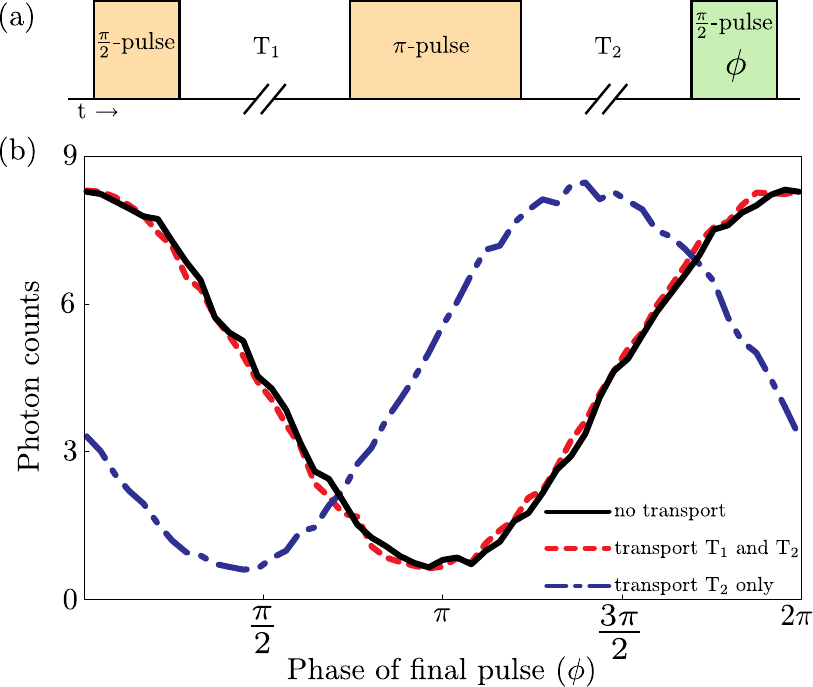}
\caption{\label{fig:deco} (a) Pulse sequence for the Ramsey experiment showing
the two $\frac{\pi}{2}$-pulses with free-precession periods $\mathrm{T}_1 =
\mathrm{T}_2$ and a spin-echo $\pi$-pulse. (b) Fluorescence versus the final
$\frac{\pi}{2}$-pulse phase $\phi$ for three cases where a single full
$\mathcal{E}$-$\mathcal{C}$-$\mathcal{E}$ transport was inserted during both
$\mathrm{T}_1$ and $\mathrm{T}_2$, during $\mathrm{T}_2$ only, or not at all.}
\end{figure}

For use in QIP, transport through a junction should preserve qubit coherence.  To
check this, we implemented a Ramsey experiment (Fig.~\ref{fig:deco}) interleaved
with junction transport.  The ion was first prepared at $\mathcal{E}$ in the
ground hyperfine state $\left| F=2, m_F=-2 \right>$, which can be detected by
fluorescence on a cycling transition. Coherent resonant Raman
$\frac{\pi}{2}$-pulses transferred the ion into an equal superposition of the
$\left| 2, -2 \right>$ (fluorescing) and $\left| 1, -1 \right>$ (nonfluorescing)
state. This was followed by a free-precession period $\mathrm{T}_1 = 280~\mu$s, a
spin-echo $\pi$-pulse (duration 1~$\mu$s), and a second free-precession period
$\mathrm{T}_2 = \mathrm{T}_1$.  Finally, the phase $\phi$ of a second
$\frac{\pi}{2}$-pulse was varied and fluorescence was collected to generate a
curve whose contrast is proportional to the ion's final coherence.  The effect of
transport on coherence was tested by inserting a single full
$\mathcal{E}$-$\mathcal{C}$-$\mathcal{E}$ transport during both $\mathrm{T}_1$
and $\mathrm{T}_2$, during $\mathrm{T}_2$ only, or not at all. The contrast when
not transporting was only $85.0\pm0.5$~\% due to the length of the experiment.
However, the contrast for one and two transports was $86.0\pm0.5$~\% and
$85.6\pm0.5$~\%, respectively, implying that transport does not affect coherence
within the measurement errors. A difference in the $B$-field magnitude
($\frac{\Delta B}{B} \approx 0.4$\%) over the $890$~nm path from $\mathcal{E}$
to $\mathcal{C}$, which results in a variable transition frequency, induced a phase
shift when only one transport was inserted, but the spin-echo effectively
canceled this phase when the transport was performed twice.

To obtain the results in Table~\ref{tab:heatingrates}, careful attention was
paid to sources of heating. Noise on the RF pseudopotential can cause ion
heating~\cite{bible,savard1997a,turchette2000a}, but this is negligible for an
ion on the RF electric-field null. However, similar heating resurfaces when the
ion is placed in a pseudopotential gradient~\cite{bible}, which is typical for
junction transport~\cite{wesenberg2008a} and can be an issue for
surface-electrode traps~\cite{chiaverini2005c}. We examine the case of an axial
pseudopotential gradient, as this caused heating along the pseudopotential
barriers near $\mathcal{C}$.

Consider an RF trap drive with a voltage noise component at $\Omega_{\textrm{RF}}
+ \omega$, where $\Omega_{\textrm{RF}}$ is the drive frequency and $\left| \omega
\right| \ll \Omega_{\textrm{RF}}$.  The trapping electric field will have the
form \begin{equation}\label{efield}
\vec{E}_{\textrm{RF}}(x,y,z,t)=\vec{E}_{0}(x,y,z) \left[ \cos{
\Omega_{\textrm{RF}} t }+\xi_{\textrm{N}} \cos{(\Omega_{\textrm{RF}} + \omega)t}
\right],
\end{equation}
where $\xi_{\textrm{N}} \ll 1$ is the noise amplitude relative to $\vec{E}_{0}$,
the ideal field. For $\xi_{\textrm{N}} = 0$, the above equation leads to the
usual pseudopotential energy
\begin{equation}\label{pseudopot}
\Phi_p = \frac{q^2}{2 m \Omega^2_{\textrm{RF}}} \left<
\vec{E}^2_{\textrm{RF}}(x,y,z,t) \right>,
\end{equation}
where $m$ and $q$ are the ion's mass and charge and the time average is performed
over the period $2\pi/\Omega_{\textrm{RF}}$. The addition of the noise term,
which beats with the carrier, introduces a noise force on the ion at frequency
$\omega$. If we take the example of $\mathcal{E}$-$\mathcal{C}$ transport
(restricting the ion to the $\hat{z}$ axis, where the field points along
$\hat{z}$), $\vec{E}_{0}(x,y,z)$ simplifies to $E_{0}(z)\hat{z}$. This yields
a noise force ($-\vec{\nabla} \Phi_p$) on the ion given by
\begin{equation}\label{noiseforce}
\vec{F}_{\textrm{N}} \approx -\frac{q^2}{2 m \Omega^2_{\textrm{RF}}} \left(
\frac{\partial}{\partial z} E^2_{0}(z) \right) \xi_{\textrm{N}} \cos{ \omega t
}~\hat{z}.
\end{equation}
Equation~(\ref{noiseforce}) can be used to relate broad-spectrum voltage noise,
given by a voltage spectral density $S_{V_{\textrm{N}}}$, to force noise
spectral density $S_{F_{\textrm{N}}}$, by use of $S_{V_{\textrm{N}}}/V_0^2 =
S_{E_{\textrm{N}}}/E_{0}^2$, where $V_0$ is the peak RF potential applied to
the trap and $S_{E_{\textrm{N}}}$ is the noise spectral density of electric
field fluctuations $E_{\textrm{N}} = \xi_{\textrm{N}} E_{0} \cos{
(\Omega_{\textrm{RF}} + \omega)t }$. For $\omega$ near the axial-mode
frequency $\omega_z$, $S_{F_{\textrm{N}}}$ heats the mode according to the rate
$\dot{\bar{n}}=S_{F_{\textrm{N}}}(\omega_z)/4 m \hbar
\omega_z$~\cite{turchette2000a}. Therefore, the heating rate for noise applied
around $\Omega_{\textrm{RF}} + \omega_z$ is
\begin{equation}\label{heatingrate}
\dot{\bar{n}} = \frac{q^4}{16
m^3 \Omega^4_{\textrm{RF}} \hbar \omega_z} \left[ \frac{\partial}{\partial z}
E^2_{0}(z) \right]^2
\frac{S_{V_{\textrm{N}}} (\Omega_{\textrm{RF}} + \omega_z)}{V_0^2}.
\end{equation}

To test this, the ion was moved to a particular location along the axis between
$\mathcal{E}$ and $\mathcal{C}$ and held there for a variable amount of time
($\geq~1$~ms) while we injected various amounts of band-limited white noise (flat
to 1~dB over 150~kHz) centered at the lower sideband $\Omega_{\textrm{RF}} -
\left| \omega_z \right|$ onto the trap RF drive.  The ion was returned to
$\mathcal{E}$ and its temperature increase was measured by use of the recooling
method.  Figure~\ref{fig:rfheating} plots the ratio of measured heating rate to
injected~$S_{V_{\textrm{N}}}$ and theoretical values of this ratio according to
Eq.~(\ref{heatingrate}) based on simulations of trap potentials, for the ion held
at several positions between $\mathcal{E}$ and $\mathcal{C}$.  The deviation
between the measured and predicted heating seems not unreasonable given the large
number of experimental variables involved that are difficult to measure directly.

We limited this heating during transport through extensive RF-filtering and by
moving the ion over the barrier as fast as possible. From heating measurements
without added noise, we estimate the ambient $S_{V_{\textrm{N}}}
(\Omega_{\textrm{RF}} \pm \omega_z)$ to be $-177$~dBc. For the transport
durations used for the experiments in Table~\ref{tab:heatingrates}, this
$S_{V_{\textrm{N}}}$ should impart $\sim$0.1~to 0.5~quanta per pass over an RF
barrier.

The other main source of energy gain was due to the digital-to-analog converters
(DACs) that supply the waveforms. While transporting through a region far from
the junction and using a fixed axial frequency (3.6~MHz), we observed energy gain
when the DAC update rate ($0.3 - 0.5$~MHz) was commensurate with the motional
frequency.  When varying the update frequency, this energy gain exhibited a
resonance at several submultiples of motional frequency with a bandwidth that
decreased when the waveforms were stretched over more updates, consequently
increasing the transport duration. This is as expected from coherent excitation.
Use of an incommensurate update rate minimized this energy gain. However,
minimizing the RF-noise heating required fast transport which broadened the
resonances.  This caused them to overlap, resulting in unavoidable energy gain.
The effect was further compounded by the changing axial frequency during full
junction transport and we experimented to find an update rate that minimized the
total energy gain. Faster DACs and strong filtering of the potentials should
alleviate this issue.

In summary, we have demonstrated reliable ion transport through a 2-D trap
junction with low motional-energy gain while maintaining internal-state qubit
coherence.  At these energies, sympathetic cooling could  be employed for
recooling in a reasonable period (for our current parameters, cooling from
motional state $n=5$ back to $n=0$ would require $100~\mu$s), while some quantum
gates are robust against small energy
gain~\cite{molmer1999a,milburn2000a,solano1999a,ospelkaus2008a} and may require
only minimal recooling.  Including recooling, the transport duration is
comparable to other QIP processes for our system, and junction transport should
not significantly slow QIP algorithms. Thus, this work demonstrates the viability
of trap arrays incorporating junctions for use in large-scale QIP using trapped
ions.

We thank J.J. Bollinger and Y. Colombe for helpful comments on the manuscript
and acknowledge the support of IARPA, ONR, and the NIST Quantum Information
program. This Letter is a contribution of NIST and not subject to U.S. copyright.


\begin{thebibliography}{27}
\expandafter\ifx\csname natexlab\endcsname\relax\def\natexlab#1{#1}\fi
\expandafter\ifx\csname bibnamefont\endcsname\relax
  \def\bibnamefont#1{#1}\fi
\expandafter\ifx\csname bibfnamefont\endcsname\relax
  \def\bibfnamefont#1{#1}\fi
\expandafter\ifx\csname citenamefont\endcsname\relax
  \def\citenamefont#1{#1}\fi
\expandafter\ifx\csname url\endcsname\relax
  \def\url#1{\texttt{#1}}\fi
\expandafter\ifx\csname urlprefix\endcsname\relax\def\urlprefix{URL }\fi
\providecommand{\bibinfo}[2]{#2}
\providecommand{\eprint}[2][]{\url{#2}}

\bibitem[{\citenamefont{Blatt and Wineland}(2008)}]{blatt2008a}
\bibinfo{author}{\bibfnamefont{R.}~\bibnamefont{Blatt}} \bibnamefont{and}
  \bibinfo{author}{\bibfnamefont{D.~J.} \bibnamefont{Wineland}},
  \bibinfo{journal}{Nature} \textbf{\bibinfo{volume}{453}},
  \bibinfo{pages}{1008} (\bibinfo{year}{2008}).

\bibitem[{\citenamefont{Monroe and Lukin}(2008)}]{monroe2008a}
\bibinfo{author}{\bibfnamefont{C.}~\bibnamefont{Monroe}} \bibnamefont{and}
  \bibinfo{author}{\bibfnamefont{M.~D.} \bibnamefont{Lukin}},
  \bibinfo{journal}{Physics World} \textbf{\bibinfo{volume}{21}},
  \bibinfo{pages}{32} (\bibinfo{year}{2008}).

\bibitem[{\citenamefont{H\"affner et~al.}(2008)\citenamefont{H\"affner, Roos,
  and Blatt}}]{haffner2008a}
\bibinfo{author}{\bibfnamefont{H.}~\bibnamefont{H\"affner}},
  \bibinfo{author}{\bibfnamefont{C.}~\bibnamefont{Roos}}, \bibnamefont{and}
  \bibinfo{author}{\bibfnamefont{R.}~\bibnamefont{Blatt}},
  \bibinfo{journal}{Phys. Rep.} \textbf{\bibinfo{volume}{469}},
  \bibinfo{pages}{155} (\bibinfo{year}{2008}).

\bibitem[{\citenamefont{Cirac and Zoller}(1995)}]{cirac1995a}
\bibinfo{author}{\bibfnamefont{J.~I.} \bibnamefont{Cirac}} \bibnamefont{and}
  \bibinfo{author}{\bibfnamefont{P.}~\bibnamefont{Zoller}},
  \bibinfo{journal}{Phys. Rev. Lett.} \textbf{\bibinfo{volume}{74}},
  \bibinfo{pages}{4091} (\bibinfo{year}{1995}).

\bibitem[{\citenamefont{Benhelm et~al.}(2008)\citenamefont{Benhelm, Kirchmair,
  Roos, and Blatt}}]{benhelm2008a}
\bibinfo{author}{\bibfnamefont{J.}~\bibnamefont{Benhelm}}
  \bibnamefont{\emph{et~al.}},
  \bibinfo{journal}{Nature Phys.} \textbf{\bibinfo{volume}{4}},
  \bibinfo{pages}{463} (\bibinfo{year}{2008}).

\bibitem[{\citenamefont{Wineland et~al.}(1998)\citenamefont{Wineland, Monroe,
  Itano, Leibfried, King, and Meekhof}}]{bible}
\bibinfo{author}{\bibfnamefont{D.~J.} \bibnamefont{Wineland}}
  \bibnamefont{\emph{et~al.}},
  \bibinfo{journal}{J. Res. Natl. Inst. Stand. Technol.}
  \textbf{\bibinfo{volume}{103}}, \bibinfo{pages}{259} (\bibinfo{year}{1998}).

\bibitem[{\citenamefont{Cirac and Zoller}(2000)}]{cirac2000a}
\bibinfo{author}{\bibfnamefont{J.~I.} \bibnamefont{Cirac}} \bibnamefont{and}
  \bibinfo{author}{\bibfnamefont{P.}~\bibnamefont{Zoller}},
  \bibinfo{journal}{Nature} \textbf{\bibinfo{volume}{404}},
  \bibinfo{pages}{579} (\bibinfo{year}{2000}).

\bibitem[{\citenamefont{Kielpinski et~al.}(2002)\citenamefont{Kielpinski,
  Monroe, and Wineland}}]{kielpinski2002a}
\bibinfo{author}{\bibfnamefont{D.}~\bibnamefont{Kielpinski}},
  \bibinfo{author}{\bibfnamefont{C.}~\bibnamefont{Monroe}}, \bibnamefont{and}
  \bibinfo{author}{\bibfnamefont{D.~J.} \bibnamefont{Wineland}},
  \bibinfo{journal}{Nature} \textbf{\bibinfo{volume}{417}},
  \bibinfo{pages}{709} (\bibinfo{year}{2002}).

\bibitem[{\citenamefont{DeVoe}(1998)}]{devoe1998a}
\bibinfo{author}{\bibfnamefont{R.~G.} \bibnamefont{DeVoe}},
  \bibinfo{journal}{Phys. Rev. A} \textbf{\bibinfo{volume}{58}},
  \bibinfo{pages}{910} (\bibinfo{year}{1998}).

\bibitem[{\citenamefont{Moehring et~al.}(2007)\citenamefont{Moehring, Madsen,
  Younge, Kohn~Jr, Maunz, Duan, Monroe, and Blinov}}]{moehring2007b}
\bibinfo{author}{\bibfnamefont{D.}~\bibnamefont{Moehring}}
  \bibnamefont{\emph{et~al.}},
  \bibinfo{journal}{J. Opt. Soc. Am. B} \textbf{\bibinfo{volume}{B24}},
  \bibinfo{pages}{300} (\bibinfo{year}{2007}).

\bibitem[{\citenamefont{Rowe et~al.}(2002)\citenamefont{Rowe, Ben-Kish,
  Demarco, Leibfried, Meyer, Beall, Britton, Hughes, Itano, Jelenkovic
  et~al.}}]{rowe2002a}
\bibinfo{author}{\bibfnamefont{M.~A.} \bibnamefont{Rowe}}
  \bibnamefont{\emph{et~al.}}, \bibinfo{journal}{Quant. Inf. Comput.}
  \textbf{\bibinfo{volume}{2}}, \bibinfo{pages}{257} (\bibinfo{year}{2002}).

\bibitem[{\citenamefont{Barrett et~al.}(2004)\citenamefont{Barrett, Chiaverini,
  Schaetz, Britton, Itano, Jost, Knill, Langer, Leibfried, Ozeri
  et~al.}}]{barrett2004a}
\bibinfo{author}{\bibfnamefont{M.~D.} \bibnamefont{Barrett}}
  \bibnamefont{\emph{et~al.}}, \bibinfo{journal}{Nature}
  \textbf{\bibinfo{volume}{429}}, \bibinfo{pages}{737} (\bibinfo{year}{2004}).

\bibitem[{\citenamefont{Hensinger et~al.}(2006)\citenamefont{Hensinger,
  Olmschenk, Stick, Hucul, Yeo, Acton, Deslauriers, Monroe, and
  Rabchuk}}]{hensinger2006a}
\bibinfo{author}{\bibfnamefont{W.~K.} \bibnamefont{Hensinger}}
  \bibnamefont{\emph{et~al.}},
  \bibinfo{journal}{Appl. Phys. Lett.} \textbf{\bibinfo{volume}{88}},
  \bibinfo{pages}{034101} (\bibinfo{year}{2006}).

\bibitem[{\citenamefont{Stick et~al.}(2006)\citenamefont{Stick, Hensinger,
  Olmschenk, Madsen, Schwab, and Monroe}}]{stick2006a}
\bibinfo{author}{\bibfnamefont{D.}~\bibnamefont{Stick}}
  \bibnamefont{\emph{et~al.}},
  \bibinfo{journal}{Nature Phys.} \textbf{\bibinfo{volume}{2}},
  \bibinfo{pages}{36} (\bibinfo{year}{2006}).

\bibitem[{\citenamefont{Huber et~al.}(2008)\citenamefont{Huber, Deuschle,
  Schnitzler, Reichle, Singer, and Schmidt-Kaler}}]{huber2008a}
\bibinfo{author}{\bibfnamefont{G.}~\bibnamefont{Huber}}
  \bibnamefont{\emph{et~al.}},
  \bibinfo{journal}{New J. Phys.} \textbf{\bibinfo{volume}{10}},
  \bibinfo{pages}{013004} (\bibinfo{year}{2008}).

\bibitem[{\citenamefont{Epstein et~al.}(2007)\citenamefont{Epstein, Seidelin,
  Leibfried, Wesenberg, Bollinger, Amini, Blakestad, Britton, Home, Itano
  et~al.}}]{epstein2007b}
\bibinfo{author}{\bibfnamefont{R.~J.} \bibnamefont{Epstein}}
  \bibnamefont{\emph{et~al.}}, \bibinfo{journal}{Phys. Rev. A}
  \textbf{\bibinfo{volume}{76}}, \bibinfo{pages}{033411}
  (\bibinfo{year}{2007}).

\bibitem[{\citenamefont{Wesenberg et~al.}(2007)\citenamefont{Wesenberg,
  Epstein, Leibfried, Blakestad, Britton, Home, Itano, Jost, Knill, Langer
  et~al.}}]{wesenberg2007a}
\bibinfo{author}{\bibfnamefont{J.~H.} \bibnamefont{Wesenberg}}
  \bibnamefont{\emph{et~al.}}, \bibinfo{journal}{Phys. Rev. A}
  \textbf{\bibinfo{volume}{76}}, \bibinfo{pages}{053416}
  (\bibinfo{year}{2007}).

\bibitem[{\citenamefont{Meekhof et~al.}(1996)\citenamefont{Meekhof, Monroe,
  King, Itano, and Wineland}}]{meekhof1996a}
\bibinfo{author}{\bibfnamefont{D.~M.} \bibnamefont{Meekhof}}
  \bibnamefont{\emph{et~al.}},
  \bibinfo{journal}{Phys. Rev. Lett.} \textbf{\bibinfo{volume}{76}},
  \bibinfo{pages}{1796} (\bibinfo{year}{1996}).

\bibitem[{\citenamefont{Walther}(1993)}]{walther1993b}
\bibinfo{author}{\bibfnamefont{H.}~\bibnamefont{Walther}},
  \bibinfo{journal}{Adv. At. Mol. Phys.} \textbf{\bibinfo{volume}{31}},
  \bibinfo{pages}{137} (\bibinfo{year}{1993}).

\bibitem[{\citenamefont{Savard et~al.}(1997)\citenamefont{Savard, OHara, and
  Thomas}}]{savard1997a}
\bibinfo{author}{\bibfnamefont{T.~A.} \bibnamefont{Savard}},
  \bibinfo{author}{\bibfnamefont{K.~M.} \bibnamefont{O'Hara}}, \bibnamefont{and}
  \bibinfo{author}{\bibfnamefont{J.~E.} \bibnamefont{Thomas}},
  \bibinfo{journal}{Phys. Rev. A} \textbf{\bibinfo{volume}{56}},
  \bibinfo{pages}{R1095} (\bibinfo{year}{1997}).

\bibitem[{\citenamefont{Turchette et~al.}(2000)\citenamefont{Turchette,
  Kielpinski, King, Leibfried, Meekhof, Myatt, Rowe, Sackett, Wood, Itano
  et~al.}}]{turchette2000a}
\bibinfo{author}{\bibfnamefont{Q.~A.} \bibnamefont{Turchette}}
  \bibnamefont{\emph{et~al.}}, \bibinfo{journal}{Phys. Rev. A}
  \textbf{\bibinfo{volume}{61}}, \bibinfo{pages}{063418}
  (\bibinfo{year}{2000}).

\bibitem[{\citenamefont{Wesenberg}(2008)}]{wesenberg2008a}
\bibinfo{author}{\bibfnamefont{J.~H.} \bibnamefont{Wesenberg}},
  \bibinfo{journal}{Phys. Rev. A}
  \textbf{\bibinfo{volume}{79}}, \bibinfo{pages}{013416}
  (\bibinfo{year}{2009}).

\bibitem[{\citenamefont{Chiaverini et~al.}(2005)\citenamefont{Chiaverini,
  Blakestad, Britton, Jost, Langer, Leibfried, Ozeri, and
  Wineland}}]{chiaverini2005c}
\bibinfo{author}{\bibfnamefont{J.}~\bibnamefont{Chiaverini}}
  \bibnamefont{\emph{et~al.}},
  \bibinfo{journal}{Quant. Inf. Comput.} \textbf{\bibinfo{volume}{5}},
  \bibinfo{pages}{419} (\bibinfo{year}{2005}).

\bibitem[{\citenamefont{M{\o}lmer and S{\o}rensen}(1999)}]{molmer1999a}
\bibinfo{author}{\bibfnamefont{K.}~\bibnamefont{M{\o}lmer}} \bibnamefont{and}
  \bibinfo{author}{\bibfnamefont{A.}~\bibnamefont{S{\o}rensen}},
  \bibinfo{journal}{Phys. Rev. Lett.} \textbf{\bibinfo{volume}{82}},
  \bibinfo{pages}{1835} (\bibinfo{year}{1999}).

\bibitem[{\citenamefont{Milburn et~al.}(2000)\citenamefont{Milburn, Schneider,
  and James}}]{milburn2000a}
\bibinfo{author}{\bibfnamefont{G.}~\bibnamefont{Milburn}},
  \bibinfo{author}{\bibfnamefont{S.}~\bibnamefont{Schneider}},
  \bibnamefont{and} \bibinfo{author}{\bibfnamefont{D.}~\bibnamefont{James}},
  \bibinfo{journal}{Fortschr. Physik} \textbf{\bibinfo{volume}{48}},
  \bibinfo{pages}{801} (\bibinfo{year}{2000}).

\bibitem[{\citenamefont{Solano et~al.}(1999)\citenamefont{Solano,
  de~Matos~Filho, and Zagury}}]{solano1999a}
\bibinfo{author}{\bibfnamefont{E.}~\bibnamefont{Solano}},
  \bibinfo{author}{\bibfnamefont{R.~L.}~\bibnamefont{de~Matos~Filho}},
  \bibnamefont{and} \bibinfo{author}{\bibfnamefont{N.}~\bibnamefont{Zagury}},
  \bibinfo{journal}{Phys. Rev. A} \textbf{\bibinfo{volume}{59}},
  \bibinfo{pages}{R2539} (\bibinfo{year}{1999}).

\bibitem[{\citenamefont{Ospelkaus et~al.}(2008)\citenamefont{Ospelkaus, Langer,
  Amini, Brown, Leibfried, and Wineland}}]{ospelkaus2008a}
\bibinfo{author}{\bibfnamefont{C.}~\bibnamefont{Ospelkaus}}
  \bibnamefont{\emph{et~al.}},
  \bibinfo{journal}{Phys. Rev. Lett.} \textbf{\bibinfo{volume}{101}},
  \bibinfo{eid}{090502} (\bibinfo{year}{2008}).

\end{thebibliography}
\end{document}